\documentclass[12pt]{elsarticle}

\usepackage{amssymb}
\usepackage{amsmath}
\usepackage{xspace}
\usepackage{xcolor}
\usepackage{hyperref}
\usepackage[utf8]{inputenc}
\usepackage{microtype}
\usepackage{qtree}
\usepackage{xpatch,letltxmacro}
\usepackage{array}
\usepackage{siunitx}
\usepackage{float}
\usepackage{listings}
\usepackage{numprint}

\usepackage[T1]{fontenc}

\journal{Astronomy and Computing}

\begin{document}

\newcommand{\stc}[0]{\texttt{sympy2c}\xspace}
\newcommand{\sympy}[0]{\texttt{SymPy}\xspace}
\newcommand{\Cpp}[0]{C/C++\xspace}
\newcommand{\PyCosmo}[0]{\texttt{PyCosmo}\xspace}

\newcommand{\bm}[1]{\textcolor{red}{BM: #1}}

\begin{frontmatter}



\title{
%\stc: generating fast C/C++ functions and ODE solvers from symbolic expressions
%\stc: fast C/C++ from symbolic expressions in Python
%\stc: fast Python extension modules from symbolic expressions
%\stc: fast Python modules in C/C++ from symbolic equations
%\stc: Fast C/C++ functions and ODE solvers from symbolic expressions in Python
\stc: from symbolic expressions to fast C/C++ functions and ODE solvers in Python
%\stc: From symbolic equations to fast C/C++ functions and ODE solvers in Python
}

\affiliation[sis]{organization={Scientific IT Services, ETH Zurich},
            addressline={Binzmühlestrasse 120}, 
            city={Zurich},
            postcode={CH-8092}, 
            country={Switzerland}}

\affiliation[cosmology_group]{organization={Institute for Particle Physics and Astrophysics, ETH Zurich},%Department and Organization
            addressline={Wolfgang-Pauli-Strasse 27}, 
            city={Zurich},
            postcode={CH-8093}, 
            country={Switzerland}}

\author[sis,cosmology_group]{Uwe Schmitt}
\ead{uwe.schmitt@id.ethz.ch}
\author[cosmology_group]{Beatrice Moser}
\author[cosmology_group]{Christiane S. Lorenz}
\author[cosmology_group]{Alexandre Refregier}

\begin{abstract}

Computer algebra systems play an important role in science as they facilitate the development of new theoretical models. The resulting symbolic equations are often implemented in a compiled programming language in order to provide fast and portable codes for practical applications. 
We describe \stc, a new Python package designed to bridge the gap between the symbolic development and the numerical implementation of a theoretical model. 
\stc translates symbolic equations implemented in the \sympy Python package to \Cpp code that is optimized using symbolic transformations. The resulting functions can be conveniently used as an extension module in Python. 
\stc is used within the \PyCosmo Python package to solve the Einstein-Boltzmann equations, a large system of ODEs describing the evolution of linear perturbations in the Universe. After reviewing the functionalities and usage of \stc, we describe its implementation and optimization strategies. 
This includes, in particular, a novel approach to generate optimized ODE solvers making use of the sparsity of the symbolic Jacobian matrix.
We demonstrate its performance using the Einstein-Boltzmann equations as a test case. \stc is widely applicable and may prove useful for various areas of computational physics. \stc is publicly available at \url{https://cosmology.ethz.ch/research/software-lab/sympy2c.html} .
\end{abstract}

\end{frontmatter}

\section{Introduction}
\label{sec:introduction}
Computer Algebra Systems (CAS), such as \texttt{Mathematica}~\cite{Mathematica} 
and \sympy~\cite{sympy}, play an important role in
scientific disciplines such as mathematics and theoretical physics, as they facilitate the development of new or modified theories. %and models.
Most often, the resulting  equations are implemented in a compiled programming language in order to provide fast, robust and portable codes for practical applications. 

In this paper, we describe \stc, a new Python package designed to bridge the gap between the symbolic development and the numerical implementation of a theoretical model. For this purpose, \stc translates symbolic equations implemented within the Python CAS \sympy to fast \Cpp code that can then be used
from Python as an extension module. 
%We explain this mechanism later in more detail.

The development of \stc started in the field of computational cosmology as part of the
Python package \PyCosmo\footnote{\url{https://cosmology.ethz.ch/research/software-lab/PyCosmo.html}}~\cite{Moser:2021rej,refregier2017pycosmo,tarsitano2020predicting}. The concept of the just-in-time compiler \texttt{HOPE}~\cite{hope} preceded the development of \stc. Among other features, \PyCosmo offers a fast solver for Einstein-Boltzmann equations, a large system of ODEs which describes the evolution of linear perturbations in the Universe~\cite{Ma:1995ey,Dodelson:2003ft}.
To improve the code structure of \PyCosmo and to make the code generator available to a
wider audience, we separated the code creation part from the other functionalities of 
\PyCosmo and thus created the separate Python package \stc.

\stc extends the basic \Cpp code generation functionalities of \sympy, for example by supporting  special functions, numerical integration, interpolation and numerical solution of ODEs. We particularly optimised the \stc code generator for high dimensional stiff ODEs  with a sparse Jacobian matrix. A direct solver for general sparse linear systems is at the heart of the ODE solver and will be described below in detail. \stc is publicly available at \url{https://cosmology.ethz.ch/research/software-lab/sympy2c.html}.

This paper is organised as follows. In Section~\ref{section:python}, we introduce the role of Python in
scientific programming, performance related aspects and the role of \stc within this context. In Section~\ref{section:stcoverview}, we give an overview of
functionalities offered by \stc and how to use the library. We list resources to access  the \stc package, its source code and documentation in Section~\ref{section:usage}.
 In Section~\ref{section:stcdetails}, we discuss the implementation and optimization details of \stc.
In Section~\ref{section:performance}, we demonstrate the performance of \stc. In Section~\ref{section:conclusions}, we summarise our conclusions.

\section{Fast Scientific Computation with Python}
\label{section:python}

%\subsection{The Python Programming Language}
Python is an interpreted, high-level, general-purpose programming language with a focus on readability and efficient programming. Nowadays, it plays an important role in many scientific disciplines. Factors for its success in science are its permissive open source license, its extendability using  \Cpp, and the availability of high-quality and easy-to-use scientific packages. 

Fundamental packages such as \texttt{numpy} and \texttt{scipy} are featured in multidisciplinary scientific journals \cite{harris2020array,virtanen2020scipy}. Python  libraries for machine learning, such as \texttt{TensorFlow}~\cite{tensorflow_developers_2022_5898685}, \texttt{PyTorch}~\cite{NEURIPS2019_9015} and \texttt{scikit-learn}~\cite{pedregosa2011scikit} are widely used~\cite{info11040193}.

Another important package is \sympy, an open source computer algebra system (CAS) written in Python. \sympy can be used directly as a Python library, and does not implement 
%instead of inventing 
its own programming language. This allows extending \sympy in Python and using \sympy with other Python libraries, without the difficulties caused by crossing language barriers.

Python is, in particular, widely used in astrophysics. Examples include \texttt{astropy}~\cite{astropy,astropy2}, a Python package for astronomy that counts more than 5800 citations at \textit{Web of Science}\footnote{\url{https://clarivate.com/products/web-of-science/}} as of March 2022, and the data processing pipelines of the Event Horizon Telescope \cite{akiyama2019first}, the LIGO observatory \cite{PhysRevD.93.122003} and the Legacy Survey of Space and Time (LSST) \cite{juric2015lsst}. We refer the reader also to the article~\cite{Faes_2012} which provides an overview of the role of Python in astronomy and science.

%\subsection{Python Performance}

Since Python is an interpreted and dynamically typed programming language, it offers great flexibility and supports an  agile development process. This, however, also implies reduced speed and higher memory consumption during run-time, features that have an impact in many scientific applications.

To circumvent this, solutions to increase execution speed have been developed and can be classified as follows:

\begin{itemize}
\item Implementation of parts of the code in a lower level language, such as \Cpp or Rust, or binding Python to existing \Cpp code. For example, large parts of \texttt{numpy}~\cite{harris2020array} and \texttt{scipy}~\cite{virtanen2020scipy} consist of a thin Python layer on top of  BLAS~\cite{blackford2002updated} and other established numerical libraries. Notable tools to simplify bindings between Python and lower level languages include \texttt{Cython}~\cite{cython}, \texttt{swig}~\cite{swig}, \texttt{pybind11}~\cite{pybind11}, \texttt{f2py}~\cite{f2py} or \texttt{PyO3}.
\item Using just-in-time compilation as provided by \texttt{HOPE}~\cite{hope}, \texttt{numba}~\cite{lam2015numba} or \texttt{PyPy}~\cite{pypy}.
\item Automatic translation of Python code to \Cpp, as supported by
 \texttt{Pythran}~\cite{pythran}.
\end{itemize}

The \stc package fits in the third category but, in contrast to the mentioned tools, creates \Cpp code from \sympy expressions rather than from existing Python functions. The package \texttt{pyodesys}~\cite{pyodesys} follows a similar approach to generate \Cpp code for evaluating the right hand side of an ODE and its Jacobian matrix. \texttt{pyodesys}  delegates these functions to existing ODE solvers, such as \texttt{pygslodeiv2}~\cite{bjorn_2020}, which do not  take sparsity into account.  The function \texttt{autowrap} of \sympy allows compilation of expressions to different back-ends such as C or FORTRAN. However,  \texttt{autowrap} does not generate code for integrals or ODEs without an explicit symbolic solution and was thus not sufficient to be used within \PyCosmo~\cite{Moser:2021rej,refregier2017pycosmo,tarsitano2020predicting}.
\stc differs from the mentioned tools by offering a very fast ODE solver by considering sparsity in the Jacobian and by implementing routines for numerical integration and  spline interpolation.

\section{Functionalities}
\label{section:stcoverview}
In this section, we describe and demonstrate the main
functionalities of \stc . For a detailed documentation of the
\stc API we refer to the \stc online documentation, available
at \url{https://cosmo-docs.phys.ethz.ch/sympy2c/}. More details
about the inner workings of \stc follow in Section
\ref{section:stcdetails}.

\subsection{Functions}

The steps to create and use a function are as follows:
\begin{enumerate}
    \item We declare a function by providing the symbolic expression which the function should evaluate and the arguments the function takes.
    \item We declare an extension module and add this function to the module.
    \item We trigger the code generation process and the compilation of the created code.
    \item We import the compiled Python extension module.
    \item We can now call the declared C-Function as a function within this module from Python.
\end{enumerate}

\noindent We demonstrate the usage pattern of \stc in Listing \ref{listing:1} where we create and use a Python extension module with a function to compute the volume of a cylinder $V=h \pi r^2$ given its height $h$ and radius $r$:

\begin{figure}[H]
\begin{lstlisting}[caption={Example of function declaration in \stc.},label=listing:1]  
from sympy2c import symbols, Function, Module
import numpy as np

r, h = symbols("r h")

module_decl = Module()
f = Function("volume_cylinder", h * np.pi * r ** 2, r, h)
module_decl.add(f)

imported_module = module_decl.compile_and_load()
print(
  "the volume of a cylinder of radius 1 and height 2 is",
  imported_module.volume_cylinder(1.0, 2.0))
)
\end{lstlisting}
\end{figure}

\begin{itemize}
    \item Line 4: Contrary to \texttt{Mathematica}, symbols are not first-class citizens in Python, thus  we have to declare  \texttt{r} and \texttt{h}.
    \item Line 6: This declares an extension module.
    \item Line 7: We declare a function named \texttt{"volume\_cylinder"} which computes the value of \texttt{h * pi * r ** 2} and takes the arguments \texttt{r} and \texttt{h}.
    \item Line 8: We add this function to the extension module
    \item Line 10: Here we trigger code generation and compilation of the generated code. We also import the generated extension module as \texttt{imported\_module}.
    \item Line 12: Now the function named \texttt{volume\_cylinder} is available as part of the \texttt{imported\_module} module.
\end{itemize}

\subsection{Integrals}
\label{subsection:integrals}

An important feature of \stc is the creation of \Cpp code for the numerical computation of integrals. \stc offers a  function \texttt{Integral} which takes an expression, the integration variable and expressions for the lower and upper integration limits. When used, for example within a \stc \texttt{Function}, \stc creates \Cpp code to compute a numerical approximation.

This is especially useful when a closed form of the anti-derivative is unknown or not computable by \sympy. Internally, \stc calls well-established QUADPACK~\cite{piessens2012quadpack} routines available in the GNU Scientific Library (\texttt{gsl})\cite{galassi-gnuscientificlibrary-2009}. These routines also support computation of indefinite integrals.

The function \texttt{Integral} from \stc returns a symbolic function which can be used as any other expression but will later be translated to a routine for the numerical approximation of the integral. \\

\noindent Listing \ref{listing:2} shows an example of how to compute  the gaussian integral $\int_{-\infty}^0 e^{-t^2} \, dt$ with \stc:

\begin{figure}[H]
\begin{lstlisting}[caption={Numerical computation of an indefinite integral.},label=listing:2]  
from sympy2c import symbols, Function, Integral, Module
import numpy as np

t = symbols("t")

integral = Integral(exp(-(t ** 2)), t, -oo, oo) 
module_decl.add(Function("gauss", integral))  

imported_module = module_decl.compile_and_load()
gauss_integral = imported_module.gauss()
print(
  "the numerical error is", 
  abs(gauss_integral - np.sqrt(np.pi)
)
\end{lstlisting}
\end{figure}

\begin{itemize}
    \item Line 6: This defines the symbolic integral $\int_{-\infty}^\infty e^{-t^2} \, dt$. \texttt{t} is the integration variable and the symbol \texttt{oo} is used by \sympy to represent $\infty$.
    \item Line 7: We add the function \texttt{gauss} which computes the given integral and takes no arguments.
    \item Line 10: We call the function \texttt{gauss} from Python, this will now execute the numerical integration.
\end{itemize}

\subsection{Cubic Spline Interpolation}

Interpolation functions can be helpful to speed up numerical computations. \stc offers a function \texttt{InterpolationFunction1D} which will create \Cpp code to call spline interpolation from \texttt{gsl}.
Such interpolation functions can also be used within the integrand or in the limits for numerical integration as presented previously in Section~\ref{subsection:integrals} or as a term in the right hand  side of the symbolic representation of an ODE as introduced later in Section~\ref{subsection:ode}.

Listing \ref{listing:3} showcases how this is supported within \stc:

\begin{figure}[H]
\begin{lstlisting}[caption={Using interpolation functions.},label=listing:3]  
from sympy2c import (symbols, Function,
                       InterpolationFunction1D, Module)
import numpy as np

t = symbols("t")
cos_approx = InterpolationFunction1D("cos_approx") 
module_decl.add(Function("f", cos_approx(t ** 2), t)) 

imported_module = module_decl.compile_and_load()

xi = np.linspace(0, 1.0, 11)
imported_module.set_cos_approx_values(xi, np.cos(xi))  
print(
   "interpolation error at x=0.5 is",
   abs(imported_module.f(0.5) - np.cos(0.5 ** 2))
)
\end{lstlisting}
\end{figure}

\begin{itemize}
    \item Line 6: This declares an interpolation function with the given name. It can be used as any other symbolic function, e.g. \texttt{sympy.sin}. Right now this serves as a "place holder function" and the user must provide appropriate values for the actual interpolation 
    %must be provided by the user 
    of the compiled extension module (see Line 12).
    \item Line 7: We add the function \texttt{f} which uses the interpolation function and computes \texttt{cos\_approx(t ** 2)}.
    \item Line 12: After compilation and import, the module contains a function \texttt{set\_cos\_approx\_values} (this name is derived from the name we specified in Line 6) to setup the interpolation. Here we provide the values for the $x_i$ values $0.0, 0.1, \ldots, 1.0$.
    \item Line 15: We check the impact of the interpolation on the actual result.
\end{itemize}

\subsection{Ordinary differential equations}
\label{subsection:ode}

\stc also generates efficient code for the numerical solution of stiff and non-stiff ordinary differential equations using the LSODA\footnote{LSODA is a variant of LSODE (\textbf{L}ivermore \textbf{S}olver for \textbf{O}rdinary \textbf{D}ifferential \textbf{E}quations) with \textbf{A}utomatic method switching} algorithm~\cite{petzold1983automatic}. LSODA automatically switches between  the Adams-Bashford method~\cite{bashforth1883attempt} for non-stiff and the BDF\footnote{\textbf{B}ackward \textbf{D}ifferentiation \textbf{F}ormula} method~\cite{curtiss1952integration} for stiff time domains. 

We used this method to replace the existing hand-crafted BDF solver from \PyCosmo to benefit from the robust and efficient step-size control as well as from the detection of stiffness and automatic adaption of the order of the two LSODA integrators. 

To demonstrate this feature, we consider the Robertson problem~\cite{robertson1966solution,hairer1996solving}, a common example of a stiff equation describing the kinetics of an autocatalytic chemical reaction of three reactants with concentrations $y_1, y_2$ and $y_3$. This problem is often used as a test problem to compare solvers for stiff ODEs. The equations for the Robertson problem are as follows:

\begin{align*} 
\dot{y_1} &= -k_1 y_1 + k_3 y_2 y_3 \\ 
\dot{y_2} &=  k_1 y_1 - k_2 y_2 -k_3 y_2 y_3 \\ 
\dot{y_3} &=  k_2 y_2^2 ,
\end{align*}

\noindent where values for the reaction coefficients are given by $k_1 = 0.04$, $k_2 = 3 \cdot 10^7$ and $k_3 = 10^4$. \\

\noindent Listing \ref{listing:4} shows how to implement this ODE using \stc: 

\begin{figure}[H]
\setcounter{figure}{3}
\renewcommand{\figurename}{Listing}
\begin{lstlisting}
import numpy as np
from sympy_to_c import Module, OdeFast, symbols

y1, y2, y3, t = symbols("y1 y2 y3 t")  
k1, k2, k3 = 1e-4, 3e7, 1e4
y1dot = -k1 * y1 + k3 * y2 * y3
y2dot = k1 * y1 - k2 * y2 - k3 * y2 * y3
y3dot = k2 * y2 ** 2                 

module_decl = Module()                      
lhs = [y1, y2, y3]
rhs = [y1dot, y2dot, y3dot]   
module_decl.add(OdeFast("robertson", t, lhs, rhs))
imported_module = module_decl.compile_and_load()    

y_start = np.array([1.0, 0.0, 0.0])    
tvec = 0.4 * 10 ** np.arange(0, 6)
rtol = 1e-6
atol = np.array([1e-8, 1e-8, 1e-10])  

result, diagnostics = imported_module.solve_fast_robertson(  
    y_start, tvec, rtol=rtol, atol=atol
) 
print(result)
\end{lstlisting}
\begin{lstlisting}[numbers=none, frame=none]
[[1.00000000e+00 0.00000000e+00 0.00000000e+00]
 [9.99640065e-01 3.33213355e-12 1.20024984e-15]
 [9.96047827e-01 3.32015942e-12 1.31485884e-14]
 [9.60826498e-01 3.20275499e-12 1.28035090e-13]
 [6.70346206e-01 2.23448735e-12 9.17743119e-13]
 [1.83165556e-02 6.10551854e-14 1.66613769e-12]]
\end{lstlisting}
\caption{Numerical solution of the Robertson ODE using \stc.}
\label{listing:4}
\end{figure}

\begin{itemize}
    \item Lines 6--8: These are the Robertson ODE equations.
    \item Line 13: Declaration to compile the ODE solver with \stc. We provide a name for the ODE, the time variable,  a list of state variables (here named \texttt{lhs}) and the list or right-hand-side expressions of the ODE.
    \item Lines 17--19: Declaration of initial values, time grid for evaluation and tolerance settings.
    \item Lines 21--24: Finally solve the ODE and print the results.
\end{itemize}

\section{Usage}
\label{section:usage}
The \stc package is hosted on \url{https://pypi.org/project/sympy2c} and hence can be installed using \texttt{pip install sympy2c}.

The source code for \stc is publicly hosted at
\url{https://cosmo-gitlab.phys.ethz.ch/cosmo_public/sympy2c} and licensed under GPLv3. 

To reduce the package size and to avoid potential license conflicts, \stc will download and compile external C code, such as the
GNU Scientific Library~\cite{galassi-gnuscientificlibrary-2009},
during the first invocation.
The documentation is available at \url{https://cosmo-docs.phys.ethz.ch/sympy2c/} and
\url{https://cosmology.ethz.ch/research/software-lab/sympy2c.html}

\section{Implementation}
\label{section:stcdetails}
In this section, we describe the main features of the implementation  of \stc. Since the fast ODE solver is at the heart of \stc and was one of the major drivers when implementing the package, we present its optimization strategy in more detail. 

\subsection{Python Extension Modules}

As mentioned in Section~\ref{section:python}, Python can be extended using \textit{extension modules}~\cite{python-extension}. The purpose of this method is either to improve performance of critical parts of a program, or alternatively to use existing external \Cpp code. 

Instead of implementing extension modules directly, programmers can use \texttt{Cython}~\cite{cython} which is a super-set of the Python language adding support for optional type declarations for variables, function arguments and return values. The \texttt{Cython} project offers tools to translate \texttt{Cython} source code into C/C++, including the code required to interact with the Python interpreter. Furthermore,  using \texttt{Cython} to implement extension modules helps to avoid errors due to reference counting of Python objects and the  created code compiles on all major operating systems and plays well with all Python versions $\ge$ 2.6 without the need to adapt the original \texttt{Cython} source code. 

\stc uses \texttt{Cython} and its tools internally to make the generated \Cpp code accessible from Python and to create code which is independent of changes between different Python versions or operating systems and compilers.

\subsection{The fast ODE solver}

Implicit solvers for stiff ODEs require to solve a linear equation involving the Jacobian matrix of the right hand side of the ODE at each time step. These systems are often sparse with a structure that depends on the interactions of the components of the ODE. 

A commonly used tool to solve dense linear systems is the LUP decomposition~\cite{gene1996van} which factors a matrix $A$ as $A = LUP$ where $L$ is a lower triangular matrix, $U$ is an upper triangular matrix and $P$ is a permutation matrix. This decomposition takes $\mathcal{O}(n^3)$ operations for a matrix of size $n \times n$. Since $P^{-1} = P^T$ for permutation matrices, solving $A x = b$ can be performed efficiently by first solving $L z = P^T b$ and then $U x = z$. Since both $L$ and $U$ are lower (respectively upper) triangular matrices, both steps involve forward (respectively backward) substitutions only.

To solve such systems, LSODA uses LAPACK's~\cite{lapack99} routines \texttt{dgetrf} to compute the LUP decomposition of general matrices, and \texttt{dgbtrf} for banded matrices. LSODA does not support other sparse matrix structures.
Due to the run-time complexity of $\mathcal{O}(n^3)$, the LUP decomposition dominates the overall run-time of LSODA for larger systems with a non-banded Jacobian.

\stc is able to gain major speed improvements by replacing the \texttt{dgetrf} routine with a specialised and fast implementation which considers the a-priori known sparsity structure derived from the symbolic representation of the ODE.

\subsubsection{Loop unrolling in the linear solver}

As an introduction into the code generator, we ignore the permutations in the LUP decomposition, and first focus on an LU decomposition.
%% Further \stc uses the common-sub-expression \texttt{cse} from \sympy to avoid duplicate evaluations of repeating terms.

\stc unrolls loops appearing in the computation of the  LU decomposition. To illustrate the idea, we assume an identity matrix $P$ and $n=4$ for a matrix

\begin{equation}
    M =
    \begin{pmatrix}
    1 & x & 0 & 0 \\
    1 & 2 & 3 & 4 \\
    y & 0 & 1 & 1 \\
    0 & 0 & x & y
    \end{pmatrix}
\end{equation}

\noindent The first step of the LU decomposition of $M$ consists then of the following nested loops:

\begin{lstlisting}[numbers=none,language=C,frame=none]
        for (k = 0; k < n - 1; k ++)
            for (int i= k + 1; i < n; i++) {
                m[i][k] /= m[k][k];
                for (int j= k + 1; j < n; j++)
                    m[i][j] -= m[i][k] * m[k][j]
            }
\end{lstlisting}

\noindent Using dedicated variables for the matrix entries instead of arrays, and unrolling the loops  during code generation, the previous code can be transformed to

\begin{lstlisting}[numbers=none,language=C,frame=none]
        double m_0_0 = 1.0;
        double m_0_1 = x;
        ...
        m_1_0 /= m_0_0;
        m_1_1 -= m_1_0 * m_0_1;
        m_1_2 -= m_1_0 * m_0_2;
        m_1_3 -= m_1_0 * m_0_3;
        m_2_0 /= m_0_0;
        ...
        m_3_2 -= m_3_1 * m_1_2;
        m_3_3 -= m_3_1 * m_1_3;
\end{lstlisting}

\noindent Since $m_{0,2} = m_{0,3} = m_{2, 1} = m_{3,0} = m_{3,1} = 0$, \stc reduces the number of computations by generating

\begin{lstlisting}[numbers=none,language=C,frame=none]
        m_1_0 /= m_0_0;
        m_1_1 -= m_1_0 * m_0_1;
        m_2_0 /= m_0_0;
        m_2_1 -= m_2_0 * m_0_1;
        m_2_1 /= m_1_1;
        m_2_2 -= m_2_1 * m_1_2;
        m_2_3 -= m_2_1 * m_1_3;     
\end{lstlisting}

\noindent The first implementation using \texttt{for} loops and arrays required 14 integer additions, 14 integer comparisons and 31 floating point operations, whereas the last specialized implementation requires 11 floating point operations only and no integer additions or comparisons.

\subsubsection{Permutation handling}

In the above example, we ignored the $P$ matrix for permuting rows for the sake of simplicity. In practice however, permuting rows to implement partial pivoting is necessary to control numerical errors~\cite{gene1996van} and cannot be ignored.
To check if we need to swap rows, the $LUP$ solver checks for every row if the corresponding element on the diagonal has a higher magnitude than the elements in the same column below the diagonal. If this is not the case, rows are swapped. 

The mathematical formulation to detect swapping is

\begin{equation}
\exists k > i\,\,\, |m_{ki}| > |m_{ii}| .
\end{equation}
This challenges our approach, since
the values of $m_{ij}$ are updated during the LUP decomposition and cannot be efficiently  computed in advance.

We use \stc within \PyCosmo to solve the same system with varying parameters over and over again. Therefore, we implemented the following adaptive strategy to mitigate this problem:

\begin{enumerate}
    \item Try to solve the linear system $A x =b$  with the optimized solver without pivoting. The specialised solver implements the necessary checks if permutations are required and falls back to a general $LUP$ solver if required. In this case we record required permutations.
    \item Afterwards, we generate and compile more optimized solvers for $A P^T_i = b P^T_i$ for recorded permutations $P_i$.
    \item In the next run, the solver switches between the existing specialized solvers if feasible. In case changes in the parameters of the ODE require a previously not appearing permutation, we fall back to the general solver as in step 1 and record permutations.
    \item Continue with step 2.
\end{enumerate}

Our experiments to solve the Einstein-Boltzmann equations with varying parameters have shown that only very few of the described iterations are required to capture the involved permutations. This is also the case for physically very unlikely parameter combinations which potentially can arise in MCMC sampling~\cite{Foreman_Mackey_2013}.

To reduce the number of permutations required, we relaxed the check above with a configurable \textit{security factor} $C > 1$ to

\begin{equation}
\exists k > i\,\,\,  |m_{ki}| > C \,|m_{ii}|.
\end{equation}

The motivation behind this change is that we assumed that numerical issues most likely arise if the affected matrix entries differ on different orders of magnitude. We used $C = 5$ or $C = 10$ in our experiments, and observed a reduction of row swaps without significant differences between computed ODE solutions.

To further speed up the fallback LUP solver, we implemented a variant which considers an a-priori known banded structure in the symbolic representation of the linear system. This avoids looping over known zero entries but at the cost of tracking band limits during matrix updates in the LUP algorithm.
This functionality is useful in our applications, but can be switched off since it increases run-time in case the given system is not banded.

\subsubsection{Splitting}

Large systems can create \Cpp functions of several millions lines of code for the unrolled LUP solvers. This can challenge the optimizer of the compiler resulting in long compilation times and high memory consumption. Our approach to mitigate these issues is to split a linear system $M x = b$ into uncoupled separate systems using Schur-complements:

\begin{enumerate}
\item We split the matrix $M$ into blocks $A, B, C, D$ with square matrices $A$ and $D$ (which may have different sizes) and also split $x$ and $b$ accordingly:
\begin{equation}
    M x = \begin{pmatrix}
    A & B \\
    C & D
    \end{pmatrix}
    \begin{pmatrix}
    x_1 \\ x_2
    \end{pmatrix}
    =
    \begin{pmatrix}
    b_1 \\ b_2
    \end{pmatrix}
\end{equation}
\item Using block-wise Gaussian elimination, we can solve this system in two steps:
\begin{align}
(A - B D^{-1} C)\, x_1 &= b_1 - B D^{-1} b_2\\
D \, x_2 &= b_2 - C x_1
\end{align}
\end{enumerate}
The code generator can compute the involved matrices symbolically and finally create two smaller \Cpp functions instead of one larger function.
\stc can apply this idea recursively to create more and smaller functions in the generated code.

Another benefit is that the fallback general LUP solver now operates on smaller systems. The solution time can thus be significantly less affected in case arising permutations are not considered already: instead of solving a system of size $n$ having run-time $\mathcal{O}(n^3)$, splitting the system of size $n$ into two uncoupled systems of size $\frac{n}{2}$ reduces the run-time by a factor of 4.  

The drawback of this approach is reduced pivoting: in the extreme case of splitting a matrix $M$ of size $n \times n$ into $n$ uncoupled solvers, \stc would not consider pivoting at all.
We did not experience precision issues in our experiments for reasonable block sizes.

\subsection{Code generation optimizations}

Since \sympy is implemented in Python, extensive symbolic manipulations performed by \stc can slow down the code generation process. This especially applies for larger ODE systems. Another factor affecting the run-time of code involving \stc is the involved compilation of the generated \Cpp code. To improve  run-time in both  cases, \stc extensively uses disk-based caches to avoid repetitive computations such that speed is significantly improved for following executions and for smaller modifications of symbolic expressions.

To improve the performance of symbolic inversion of matrices needed in the splitting approach, we replaced the \texttt{inv} function from \sympy by recursive application of the following equation for $M$ of size $n \times n$ and quadratic matrices $A$ of size $\lfloor \frac{n}{2} \rfloor$ and
$D$ of size $\lceil \frac{n}{2} \rceil$

\begin{align}
M =
\begin{pmatrix}
   A & B \\
   C & D \\
\end{pmatrix}^{-1} &= \begin{pmatrix}
    {A} ^{-1}+ {A} ^{-1}{B} R {CA} ^{-1}
    &
    - {A} ^{-1} {B} R
    \\
    -R {CA} ^{-1}
    &
    R
\end{pmatrix} \\
R &= \left({D} - CA ^{-1} {B} \right)^{-1}.
\end{align}

\section{Performance}
\label{section:performance}

\subsection{Setup}

To test the performance of \stc, we consider the numerical solution of the Einstein-Boltzmann equations~\cite{Ma:1995ey,Dodelson:2003ft} implemented in \PyCosmo~\cite{refregier2017pycosmo,Moser:2021rej}.
This is a system of first-order linear homogeneous differential equations describing the evolution of linear perturbations in the Universe. 
The overall structure of the equation system is of the form
\begin{equation}
    \mathbf{y}'(t) = \mathbf{J}(t) \mathbf{y}(t)
\end{equation}
where $\mathbf{y}$ is a vector of perturbation fields, prime denotes derivative with respect to a time variable $t$, and $\mathbf{J}$ is a time dependent Jacobian matrix.
Generally $\mathbf{y}$ also depends on the wave number $k$, so this equation needs to be solved for a vector of $k$ values. 

These equations are crucial for accurate cosmological model predictions. The radiation fields, photons' temperature and polarization and (massless) neutrinos' temperature, need to be described as an infinite hierarchy of multipoles, which are truncated to finite sums of length $l +
1$ to allow numerical solutions.  Thus, the size of the ODE system depends on a
parameter \texttt{l\_max} which leads to a system of size $5 + 3 \times
(\texttt{l\_max} + 1)$. In practical cases, the resulting dimension of
pertubation fields $\mathbf{y}$ can be several hundreds. This large
dimensionality in addition to the stiff nature of the equations make their
fast numerical solution challenging. For a theoretical analysis of this
system of equations we refer to \cite{boltzman}.

We published time measurements for different variations of the Einstein-Boltzmann equations  in~\cite{Moser:2021rej}.
Below, we focus on demonstrating and comparing the effects 
of the discussed optimizations. The presented time measurements were performed on the high performance computing cluster \texttt{Euler} at ETH Zurich. We allocated a full node equipped with an EPYC 7742 processor from AMD. Further, we measured timings five times for every configuration and always report the fastest of these runs.

As noted above, the system depends on the wave number $k$. Higher values of $k$ increase oscillations in the solution, which enforce smaller step-sizes in LSODA, and thus lead to longer computation times.

We compare run-times of different $k \in \{ 10^{-5}, 0.1, 1, 10\} \,h{\textrm{Mpc}^{-1}}$ and  $\texttt{l\_max} \in \{10, 50, 100 \}$ resulting in equations systems of size
$n \in \{38, 158, 308 \}$.

\subsection{Time Measurements}

Tables \ref{timings:0.1}, \ref{timings:1.0} and \ref{timings:10}  show the execution time measurements  for $k=0.1, 1.0 $ and $10.0 \,h{\textrm{Mpc}^{-1}}$. The presented columns are:

\begin{itemize}
    \item \textit{n} is the size of the system.
    \item \textit{mode}: 
    \begin{itemize}
        \item \textit{full} indicates that we disabled all optimizations and LSODA always uses the fallback general LUP solver.
        \item \textit{banded} indicates that we disabled all optimizations and LSODA uses the fallback LUP solver with the optimizations for the banded structure of the Jacobian matrix as described before.
        \item \textit{optimized} refers to using the optimized linear solver using unrolled loops and avoiding computations involving known zeros.
    \end{itemize}
    \item $T_{\texttt{LUP}}$ is the total time spent in the full LUP linear solver. This solver is used when the optimized solver is disabled (modes \textit{full} and \textit{banded}), or when the optimized solver encounters an unknown row permutations and switches to the fallback LUP solver (mode \textit{optimized}). Reporting the value $0.00$ in \textit{optimized} mode indicates that the fallback solver was not required during solving the ODE system.
    \item $T_{\texttt{optim}}$ is the total time spent in the optimized linear solver.
    \item $T_{\texttt{total}}$ is the overall time required to solve the differential equation. In addition to the time spent in solving linear equations measured as $T_{\texttt{LUP}}$  resp.  $T_{\texttt{optim}}$,
    this also includes time spent in the actual LSODA algorithm.
    \item \textit{S} is the achieved speedup to solve the ODE. This  factor describes the reduction in run time compared to the baseline mode \textit{full}.
    
\end{itemize}

%%%%%%%%%%%%%%%%%%%%%%%%%%%%%%%%%%%%%%%%%%%%%%% k = 0.1

\begin{table}[t]

\caption{Timings for $k=0.1 \,h{\textrm{Mpc}^{-1}}$}

\label{timings:0.1}

\npdecimalsign{.}
\begin{center}
\begin{tabular}{|r|l|n{4}{2}|n{4}{2}|n{3}{2}|r|} 
\hline 
{ n }
& mode 
& { $T_{\texttt{LUP}}$[s] }  
& { $T_{\texttt{optim}}$[s] }
& { $T_{\texttt{total}}$[s] }
& { S } \\
\hline
38 & full & 4.53e-2 &  & 5.01e-2 & 1.00\\
 &  banded & 1.86e-2 &  & 2.32e-2 & 2.15\\
 & optimized & 0.00 & 2.16e-3 & 6.76e-3 & 7.41\\
 & & & & & \\
158 &  full & 3.28 &  & 3.29 & 1.00\\
 &   banded & 1.82 & & 1.83 & 1.80\\
 &  optimized & 0.00 & 7.30e-3 & 1.80e-2 & 183.36\\
 & & & & & \\
308 &  full & 2.65e1 &  & 2.66e1 & 1.00\\
 &   banded & 1.29e1 &  & 1.30e1 & 2.05\\
 &  optimized & 0.00 & 1.71e-2 & 3.71e-2 & 716.66\\
\hline

\end{tabular}
\end{center}

\end{table}
Tables \ref{timings:0.1}, \ref{timings:1.0} and \ref{timings:10}
show that the optimized solver spends no measurable time in the
fallback LUP solver, indicating that no row permutations
were required to ensure numerical precision. The speed-up we
achieved is almost independent of $k$ but significantly
influenced by the size $n$ of the system.

We did not find any numerical difference in the computed solutions for the different linear solvers. This is not surprising since the different solvers just reduce the number of
computations involving zeros and else run the same algebraic operations in the same order.

%%%%%%%%%%%%%%%%%%%%%%%%%%%%%%%%%%%%%%%%%%% k = 1.0

\begin{table}[H]
\npdecimalsign{.}

\caption{Timings for $k=1.0 \,h{\textrm{Mpc}^{-1}}$}
\label{timings:1.0}

\begin{center}
\begin{tabular}{|r|l|n{4}{2}|n{4}{2}|n{3}{2}|n{3}{2}|} 
\hline 
{ n }
& mode 
& { $T_{\texttt{LUP}}$[s] }  
& { $T_{\texttt{optim}}$[s] }
& { $T_{\texttt{total}}$[s] }
& { S } \\
\hline
38 & full & 4.23e-1 &  & 4.65e-1 & 1.00\\
 &  banded & 1.68e-1 &  & 2.08e-1 & 2.24\\
 &  optimized & 0.00 & 1.70e-2 & 5.14e-2 & 9.04\\
 & & & & & \\
158  & full & 2.57e1 &  & 2.58e1 & 1.00\\
 &  banded & 1.44e1 &  & 1.46e1 & 1.77\\
 &  optimized & 0.00 & 6.83e-2 & 1.65e-1 & 156.50\\
& & & & & \\
308  & full & 2.07e2 &  & 2.07e2 & 1.00\\
 &  banded & 1.05e2 &  & 1.06e2 & 1.96\\
 &  optimized & 0.00 & 1.38e-1 & 2.97e-1 & 697.21\\ 
\hline

\end{tabular}
\end{center}

\end{table}

%%%%%%%%%%%%%%%%%%%%%%%%%%%%%%%%%%%%%%%%%%%%%% k = 10

\begin{table}[H]

\caption{Timings for $k=10 \,h{\textrm{Mpc}^{-1}}$}
\label{timings:10}

\npdecimalsign{.}
\begin{center}
\begin{tabular}{|r|l|n{4}{2}|n{4}{2}|n{3}{2}|r|} 
\hline 
{ n }
& mode 
& { $T_{\texttt{LUP}}$[s] }  
& { $T_{\texttt{optim}}$[s] }
& { $T_{\texttt{total}}$[s] }
& { S } \\
\hline
38 &  full & 3.97 &  & 4.35 & 1.00\\
 &   banded & 1.63 &  & 2.00 & 2.18\\
 &  optimized & 0.00 & 1.71e-1 & 5.00e-1 & 8.70\\
  & & & & & \\
58 &  full & 2.38e2 &  & 2.39e2 & 1.00\\
 &   banded & 1.43e2 &  & 1.44e2 & 1.66\\
 &  optimized & 0.00 & 5.82e-1 & 1.37 & 175.39\\
  & & & & & \\
158 &  full & 1.99e3 &  & 1.99e3 & 1.00\\
 &   banded & 1.01e3 &  & 1.01e3 & 1.97\\
 &  optimized & 0.00 & 1.37 & 2.88 & 691.24\\
 
\hline

\end{tabular}
\end{center}

\end{table}

\subsection{Impact of the Fallback LUP solver}

To trigger the use of the fall-back solver we had to choose $k=10^{-5} \,h{\textrm{Mpc}^{-1}}$. As shown in table \ref{timings:using-fallback}, in this case the values in the column $T_\textrm{LUP}$ are non-zero in the  \textit{optimized} mode. The speed-up of the optimized solver is reduced significantly in this situation.

%%%%%%%%%%%%%%%%%%%%%%%%%%%%%%%%%%%%%%%%%% with fallback

\begin{table}[H]
\caption{Timings for $k=10^{-5} \,h{\textrm{Mpc}^{-1}}$, using the LUP fallback solver}
\label{timings:using-fallback}

\npdecimalsign{.}
\begin{center}
\begin{tabular}{|r|l|n{4}{2}|n{4}{2}|n{3}{2}|r|} 
\hline 
{ n }
& mode 
& { $T_{\texttt{LUP}}$[s] }  
& { $T_{\texttt{optim}}$[s] }
& { $T_{\texttt{total}}$[s] }
& { S } \\
\hline

38 & full & 5.22e-3 &  & 5.78e-3 & 1.00\\
 &   banded & 3.50e-3 &  & 4.40e-3 & 1.31\\
 &  optimized & 1.65e-3 & 6.24e-5 & 2.29e-3 & 2.53\\
 
  & & & & & \\
  
58 &  full & 3.47e-1 &  & 3.49e-1 & 1.00\\
 &   banded & 1.93e-1 & & 1.95e-1 & 1.79\\
 &  optimized & 1.44e-1 & 8.71e-4 & 1.47e-1 & 2.37\\
 
  & & & & & \\

158 &  full & 3.18 &  & 3.18 & 1.00\\
 &   banded & 1.15 &  & 1.16 & 2.75\\
 &  optimized& 0.82 & 1.16e-3 & 8.26e-1 & 3.85\\

\hline

\end{tabular}
\end{center}

\end{table}

Table \ref{timings:recompilation} shows the execution time after
updating and recompiling the generated \Cpp code based on the recorded row-permutations from the previous run. We can see that the optimized solver now achieves reduction in execution time similar to the measurements we presented before.

%%%%%%%%%%%%%%%%%%%%%%%%%%%%%%%%%%%%%%%%%%%%%%%% after recompile

\begin{table}[H]
\npdecimalsign{.}

\caption{Improved timings for $k=10^{-5} \,h{\textrm{Mpc}^{-1}}$ after recompilation}
\label{timings:recompilation}

\begin{center}
\begin{tabular}{|r|l|n{4}{2}|n{4}{2}|n{3}{2}|r|} 
\hline 
{ n }
& mode 
& { $T_{\texttt{LUP}}$[s] }  
& { $T_{\texttt{optim}}$[s] }
& { $T_{\texttt{total}}$[s] }
& { S } \\
\hline
38 & full & 5.22e-3 & & 5.78e-3 & 1.00\\
 &   banded & 3.50e-3 & & 4.40e-3 & 1.31\\
 &  optimized & 0.00 & 2.87e-4 & 7.35e-4 & 7.53\\
 
  & & & & & \\
  
58 &  full & 3.47e-1 & & 3.49e-1 & 1.00\\
 &   banded & 1.93e-1 & & 1.95e-1 & 1.79\\
 &  optimized & 0.00 & 1.18e-3 & 2.17e-3 & 189.31\\
 
  & & & & & \\

158 &  full & 3.18 &  & 3.18 & 1.00\\
 &   banded & 1.15 & & 1.16 & 2.75\\
 & optimized & 0.00 & 1.96e-3 & 3.58e-3 & 881.80\\

\hline
\end{tabular}
\end{center}
\end{table}

\subsection{Runtime scaling}

\begin{figure}[ht!]
    
    \centering
    \includegraphics[width=0.9\linewidth]{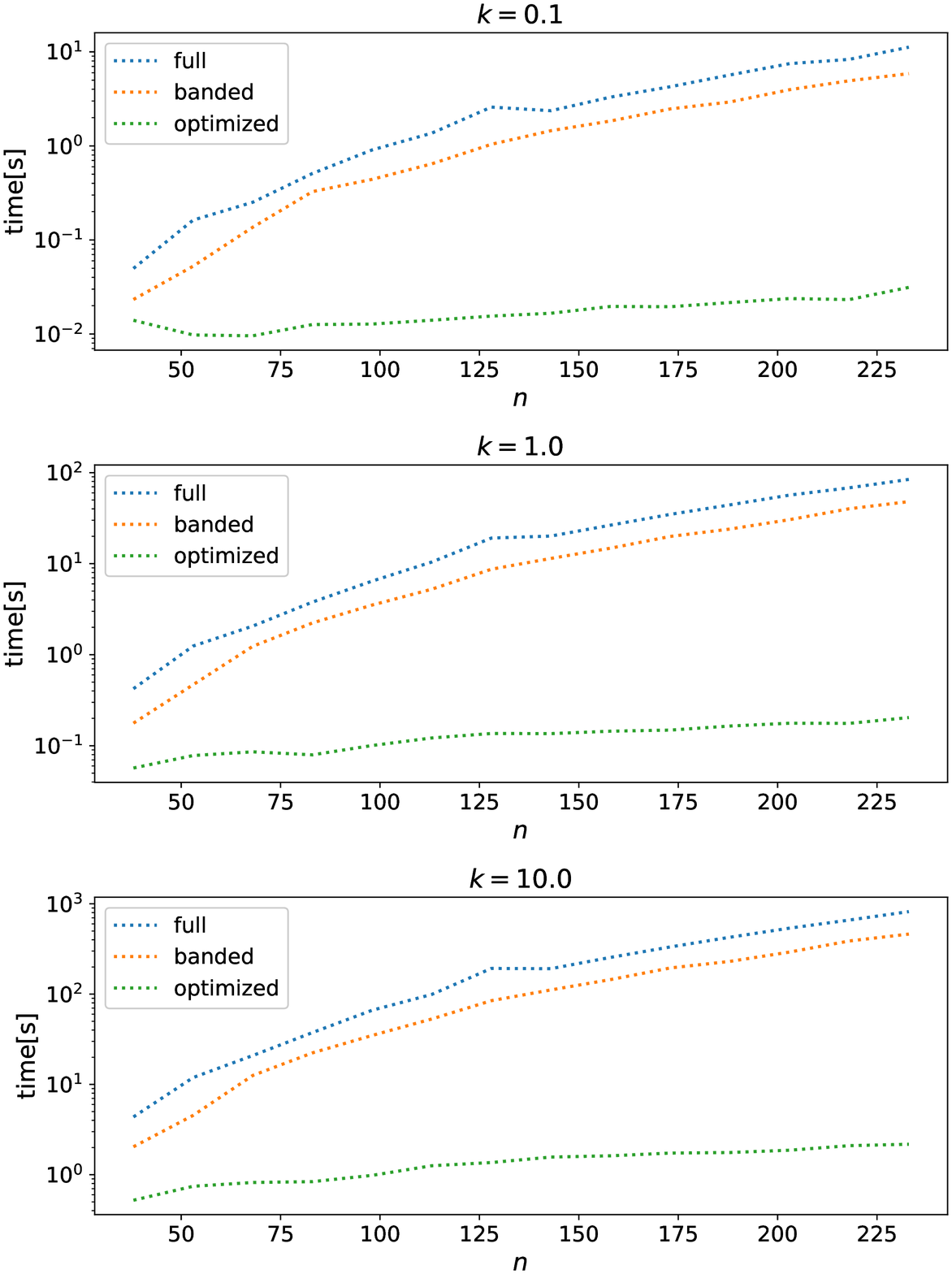}
    \caption{Run-time comparisons of the different optimization levels for different values of the wave number $k$ (in  $h{\textrm{Mpc}^{-1}}$). $n$ is the size of the ODE system and the measured time is the total execution time to compute the ODE solution}
    \label{fig:complexity}
\end{figure}

We investigated the run-time scaling of the LSODA solver for different linear solvers as a function of the number of equations $n$. Figure~\ref{fig:complexity} shows how the run-time of both variants of the LUP solver (\textit{full} and \textit{banded}) grows similarly with $n$, whereas the dependence on $n$ is flatter for the \textit{optimized} solver.

We performed polynomial fits of different orders and compared them using the AIC and BIC model selection criteria~\cite{burnhamdr} using the Python package \texttt{statsmodels}~\cite{seabold2010statsmodels}. We find that
\begin{itemize}
    \item 
    the run-times for solving systems of $n$ equations in the \textit{full} and \textit{banded} modes follow  
    \begin{equation}
T_{\textit{mode},k}(n) \approx \alpha_{\textit{mode},k} + \beta_{\textit{mode},k} \, n^3
\end{equation}
 with fitted parameters $\alpha_{\textit{mode},k} $ and $\beta_{\textit{mode},k} $. The goodness of fit  resulted in adjusted $r^2_\textrm{adj}$ values $\ge 0.99$ for all $k$ values considered.
 
    \item 
    the run-time for solving a system of $n$ equations in the \textit{optimized} mode grows linearly in $n$: \begin{equation}T_k(n) \approx \alpha_k + \beta_k \, n\end{equation}
with fitted parameters $\alpha_k$ and $\beta_k$. Fits achieved adjusted $r^2_\textrm{adj}$ values $\ge 0.98$ for $k = 1.0, 10.0 \,h{\textrm{Mpc}^{-1}}$ and $r^2_\textrm{adj} \ge0.87$ for $k=0.1\,h{\textrm{Mpc}^{-1}}$. Residuals for all $k$ appeared randomly distributed. This also indicates that the lower $r^2_\textrm{adj}$ value  for $k = 0.1\,h{\textrm{Mpc}^{-1}}$ is caused by measurement noise and that there is no remaining term growing faster than $n$ which our fit may have missed.
        \end{itemize}

Users of \stc thus benefit most from our optimized solver for large systems, but at the cost of upfront code generation and compilation times.

\section{Conclusions}
\label{section:conclusions}
We presented the new \stc Python package for generating fast \Cpp code from symbolic expressions. \stc supports the creation of functions and solvers for stiff and non-stiff ordinary differential equations. It also implements functions to support numerical interpolation and integration. \stc is general and widely applicable and may thus prove useful for various areas of computational physics.

Our run-time measurements show that the optimization of the linear solver yield a significant improvement on the overall runtime performance of the ODE solver, in particular  for larger systems. The overhead of code generation and compilation time  limits application scope of the ODE solver to situations where the same ODE has to be solved many times with varying coefficients or initial conditions. 
To mitigate this, we plan to reduce the compilation times in future versions of \stc by creating more and smaller files to support the optimization step of the underlying compiler and to enable parallel compilation of different source code files.

\section{Acknowledgements}

The authors thank Joel Mayor for useful discussions on extensions of \PyCosmo. This work was supported in part by grant No 200021\_192243 from the Swiss National Science Foundation.
\stc depends on the Python packages \texttt{Cython}~\cite{cython} and \texttt{sympy}~\cite{sympy}. Further \stc makes use of  the 
GNU Scientific Library (\texttt{gsl})\cite{galassi-gnuscientificlibrary-2009}
and the 
LSODA source code~\cite{petzold1983automatic}. Many ideas in \stc are influenced by the Python package \texttt{HOPE}~\cite{hope} and previous developments in \PyCosmo~\cite{refregier2017pycosmo,tarsitano2020predicting}.

Benchmarks were run on the  \texttt{Euler} computing cluster at ETH Zurich\footnote{\url{https://scicomp.ethz.ch}} provided by the HPC team from \textit{Scientific IT Services of ETH}\footnote{\url{https://sis.id.eth.ch}}.

\clearpage
\bibliographystyle{elsarticle-num} 
\bibliography{sympy_to_c}




% \end{thebibliography}
\end{document}